\documentclass[12pt]{article}

\pdfoutput=1
\usepackage{jheppub}
\usepackage{url}
\usepackage{color}
\usepackage{fancyhdr}
\usepackage{amsthm}
\usepackage{subfigure}
\usepackage{enumerate}

\newcommand{\be}{\begin{equation}}
\newcommand{\ee}{\end{equation}}
\newcommand{\bal}{\begin{aligned}}
\newcommand{\eal}{\end{aligned}}
\newcommand{\Log}{\mathrm{ln}}

\def\xptn#1{\langle #1 \rangle}
\def\tr{\mathop {{\rm \, Tr}} \nolimits}         
\def\Tr{\tr}

\font\openface=msbm10 at10pt
\def\Minkowski     {{\hbox{\openface M}}}

\title{Spacetime Entanglement Entropy in 1+1 Dimensions}

\author[a,b]{\small Mehdi Saravani,}
\author[a,c]{Rafael D. Sorkin,}
\author[a,b]{and Yasaman K. Yazdi}

\affiliation[a]{Perimeter Institute for Theoretical Physics, 31 Caroline St. N., Waterloo ON, N2L 2Y5, Canada}
\affiliation[b]{Department of Physics and Astronomy, University of Waterloo, Waterloo ON, N2L 3G1, Canada}
\affiliation[c]{Department of Physics, Syracuse University, Syracuse NY, 13244-1130, U.S.A.}

\emailAdd{msaravani@perimeterinstitute.ca}
\emailAdd{rsorkin@perimeterinstitute.ca}
\emailAdd{yyazdi@perimeterinstitute.ca}

\begin{document}

\abstract{%
Reference \cite{SorkinS} defines an entropy for a gaussian scalar field
$\phi$ in an arbitrary region of either a causal set or a continuous
spacetime, given only the correlator $\xptn{\phi(x)\phi(y)}$ within the
region.  As a first application, we compute numerically the entanglement
entropy in two cases where the asymptotic form is known or suspected
from conformal field theory, finding excellent agreement when the
required ultraviolet cutoff is implemented as a truncation on spacetime
mode-sums.  We also show how the symmetry of entanglement entropy
reflects the fact that $RS$ and $SR$ share the same eigenvalues, $R$ and $S$
being arbitrary matrices.
}

\keywords{}

\pagestyle{plain}
\thispagestyle{empty}


\maketitle

\flushbottom

\section{Introduction}
It is customary to conceive of entropy in a quantum field
theory as defined relative to a spacelike surface $\Sigma$ on which the
momentary state of the field is represented by a density-matrix
$\rho(\Sigma)$.  But for some purposes a more global notion of entropy
would be preferable.
For one thing, the notion of state at a moment of time might not survive
in quantum gravity, and it seems in special jeopardy in relation to
discrete theories, including causal sets \cite{Bombelli:1987aa} and others.
Moreover, even in flat spacetimes, quantum fields are believed to be too
singular to be meaningfully restricted to lower dimensional
submanifolds,
and in the context of quantum gravity with its fluctuating causal structure,
this problem can only become worse.
A more global conception of entropy is also called for if one aims at a
path-integral or ``histories-based'' formulation of quantum mechanics.
And such a conception would seem especially fitting in connection with
black holes, whose very definition is global in character.

But over and above all these considerations stands the question of an
ultraviolet ``cutoff''.  If one seeks to compute, for example, the
entropy of entanglement of a scalar field between the interior and
exterior of a black hole, one inevitably encounters a divergent answer
that traces its existence to the infinitely many high frequency modes of
the field in the neighbourhood of the horizon.  Within a particular
Cauchy surface $\Sigma$, one can cut these modes off at some given
wavelength $\lambda$ but there is no guarantee that one would obtain the
same answer if one tried to use the same cutoff with a different
hypersurface.  And without such a guarantee, it seems hard to feel fully
confident in basic results like the proportionality of entanglement
entropy to area \cite{padova,Bombelli:1986rw}.

Thus arises the need for a covariant (locally Lorentz invariant) cutoff
or --- better still --- a more fundamental theory of spacetime structure
that would furnish nature's own regularization scheme.  Based on evidence
from causal sets and such attempts as non-commutative geometry, one can
anticipate that an entropy defined this way would need to refer to whole
regions of spacetime rather than simply hypersurfaces.  (For example,
the spatio-temporal volume-element is invariant, but the spatial
volume-element is not, a basic underpinning of causal set theory.)
The need
for a covariant discreteness or other covariant cutoff thus gives rise
to a further need,
the need for
a definition of entropy
that does not rely
on the notion of state on a hypersurface.

Recently, an expression of this sort
has been derived,
which,
for a gaussian scalar field
(a free scalar field in a gaussian state),
deduces an entropy
for an arbitrary region $R$ of spacetime
from the correlation function of the field
within that region,
$\langle0|\phi(x)\phi(x')|0\rangle$,
where $|0\rangle$ is the given gaussian state \cite{SorkinS}.
When $R$ is globally hyperbolic with Cauchy surface $\Sigma$, the
resulting entropy can be identified with that of $\Sigma$,
but
unlike with previous
formalizations of the entropy concept,
this expression is {\it  covariant} in the sense that
it involves only space-time quantities.~\footnote%
{This ``covariant'' entropy agrees formally with the usual one in
  situations where both can be defined, but it applies also to
  non-globally hyperbolic spacetime regions, to causal sets, and more
  generally to any algebra with bosonic generators, as illustrated by
  the quantum theories we study herein.  (A fermionic analog also
  exists.)  The new entropy is also new in the sense that it demands a
  different sort of UV cutoff, and each different way of introducing a
  cutoff is technically a different definition of entropy.  }


An advantage of this method is that it is applicable to both continuum
spacetimes and to discrete causal sets, where the fundamental
discreteness provides naturally a frame-independent cutoff.  In this
paper we study the new
entropy-expression
in the continuum, in order
to be able to compare its behaviour with known results about entanglement
entropy arising in the context of conformal field theory (CFT).

Let us recall how entanglement entropy can be captured by the new
formula.
In conventional treatments,
entropy is identified with
the ``Gibbs entropy''
\be
  S = \Tr\rho\ln\rho^{-1}
\ee
of a density-matrix $\rho(\Sigma)$,
where $\Sigma$ is a hypersurface and $\rho$ is a ``statistical state'' for
this hypersurface.
If $\Sigma$ is divided into complementary subregions $A$ and $B$, then
the reduced density matrix for subregion $A$ is
\be
    \rho_A = \Tr_B \rho
\ee
and its entropy is
\be
   S_A = -\Tr\rho_A\ln\rho_A  \ .
\ee
Provided that the full density-matrix $\rho(\Sigma)$ is pure,
one can refer to $S_A$ (which then necessarily equals $S_B$)
as the {\it entanglement entropy} between $A$ and $B$.
In the new approach, the entropy of $A$ is taken to be that of the
spacetime region $R_A=D(A)$, where $D(A)$ is the ``causal development''
or ``domain of dependence'' of $A$.
(More generally, $R_A$ can be any sub-region of $D(A)$ that contains $A$
in its interior.)
In spacetime language, $S_A$ can be described as the entropy of
entanglement between $R$ and its so-called ``causal complement''.
And -- modulo the usual caveats about $S_A$ being infinite -- it is a
theorem that the new and old approaches produce the same result for
$S_A$.

In what follows, we describe the spacetime entropy formula more fully
and apply it to compute entanglement entropies in some cases of
interest.

In Section 2 we review the derivation of the formula and point out that
it yields correctly the thermal entropy of a simple harmonic oscillator,
as shown in detail in Appendix B.

In Section 3 we consider a two-dimensional ``causal diamond'' immersed
in the vacuum within a larger causal diamond (our choice of vacuum
being described more fully in Appendix A).  The spacetime entropy of
the smaller diamond in this case measures (when interpreted spatially)
the entanglement between a line-segment and its complement within a
larger line-segment.
In Section 4 we consider a similar
case
corresponding
spatially to a
segment embedded in a half-line.  In both situations we carry out the
computation numerically for a  massless scalar field.

Finally, we devote Appendix C to an ab initio proof of the symmetry of
entanglement entropy between a region and its complementary region.  The
pleasantly simple derivation given there relates this symmetry directly
to the fact that the product of two matrices has the same eigenvalues,
no matter in which order the matrices are multiplied.

\section{The Entropy of a Gaussian Field}
%
Let us review the derivation in \cite{SorkinS}.
We start by considering a single
``degree of freedom'' corresponding to
a conjugate pair of variables $q$ and $p$ that satisfy $[q,p]=i$.
From them we can form three independent
correlators,
$\langle qq\rangle$,
$\langle pp\rangle$,
and
$Re \langle qp\rangle$.
In a $q$-basis
for this ``degree of freedom'',
a general Gaussian density matrix
takes the form
\be
  \rho(q,q')
  \equiv\langle q |\rho|q'\rangle
  \propto
  \exp{\left[-\frac{A}{2}(q^2+q'^2)+\frac{iB}{2}(q^2-q'^2)-\frac{C}{2}(q-q')^2\right]}
  \ ,
\ee
where the real parameters $A$, $B$, and $C$ are completely determined by the
above correlators.

Given that the entropy,
$S(\rho)=\Tr\rho\ln\rho^{-1}$,
has to be dimensionless
and invariant under unitary transformations,
it can only depend on the combination:
\be
  \langle qq\rangle \langle pp\rangle-(Re\langle qp\rangle)^2=\frac{C}{2A}+\frac{1}{4}
\ee
In the original work on entanglement entropy in references \cite{padova,Bombelli:1986rw},
it was shown that $S(\rho)$ takes the form
\be
    -S = \frac {\mu\ln\mu + (1-\mu)\ln(1-\mu)} {1-\mu}
\ee
with
\be
   \mu = \frac {\sqrt{1+2C/A}-1} {\sqrt{1+2C/A}+1}
\ee
To express $S$ directly in terms of the correlators, we can
introduce the ``Wightman'' and ``Pauli-Jordan'' matrices,
\[ W=\left( \begin{array}{cc}
  \langle qq \rangle & \langle qp \rangle \\
  \langle pq \rangle & \langle pp \rangle  \\
  \end{array} \right) \]
and
\[ i\Delta=
  \left( \begin{array}{cc}
  0 & i  \\
  -i & 0  \\
  \end{array} \right) \ .
\]
The matrix $W$ corresponds in the field theory to
$W(x,x')=\langle 0|\phi(x)\phi(x')|0\rangle$,
while
$\Delta$ gives the imaginary part of $W$
and corresponds to the commutator function defined by
$i\Delta(x,x')=[\phi(x),\phi(x')]$.
Then
\be
    S=(\sigma+1/2)\ln(\sigma+1/2)-(\sigma-1/2)\ln(\sigma-1/2), \label{s2}
\ee
where $\pm i\sigma$ are the eigenvalues of $\Delta^{-1}R$,
with $R\equiv Re[W]$ being the (componentwise) real part of $W$.
We can further simplify \eqref{s2} by writing it in terms of the
eigenvalues of
 $\Delta^{-1}W=\Delta^{-1}R+i/2$ rather than those of
 $\Delta^{-1}R$.
Calling these eigenvalues $\pm i\omega_{\pm}$,
we have $\pm i\omega_{\pm}=i(1/2\pm\sigma)$,
and our formula for the entropy becomes
\be
   S = \omega_{+} \ln\omega_{+} - \omega_{-} \ln\omega_{-}  \label{s3}
\ee
where $\omega_{+}$ and $-\omega_{-}$ are now the two solutions $\lambda$ of the
generalized eigenvalue problem:
\be
  W \, v = i\lambda \; \Delta \, v  \ .                \label{gee}
\ee
%
%
In terms of these eigenvalues \eqref{s3} becomes simply
\be
  S=\sum \lambda \, \ln |\lambda| \ . \label{s4}
\ee

In the special case just treated, $\Delta$ was invertible and we could just
as easily have written \eqref{gee} as an eigenvalue equation for
$\Delta^{-1}W$.  In general, however, $\Delta$ will have ``zero modes''
and will not be invertible, which is why we wrote \eqref{gee} in
the way that we did.  When $\Delta$ is not invertible we can still
define $\lambda$ via \eqref{gee}, but in solving it we will add the
further proviso\footnote%
{Instead of restricting $v$ in this way, we could instead construe it as
 an equivalence-class of solutions, two solutions being equivalent when
 their difference is annihilated by $\Delta$.  This quotient
 construction is equivalent to limiting $v$ to the image of $\Delta$,
 but it is more ``invariant'', because $u$ and $\Delta{u}$ belong to
 different vector spaces, whence neither $W$ nor $\Delta$ can act on
 $v=\Delta{u}$ without the aid of an auxiliary metric (which for us will
 be the $L^2$ inner product).  That the two ways of proceeding yield the
 same entropy in the end can be verified explicitly in the oscillator
 example of Appendix B.}
%
%
that $v$ must belong to the {\it\/image\/} of $\Delta$.
With this proviso, our formulas for a single degree of freedom remain
valid
for many degrees of freedom,
and
$S$ is given by \eqref{s4} with the sum taken
over the full set of independent solutions of \eqref{gee}.

To fully justify this prescription
and delineate its exceptional cases
we would need to take a detour into
operator algebras and irreducible representations in Hilbert space.
This would lead to a more algebraic definition of entropy and a proof of
its equivalence to (1.1) under the assumption that an irreducible
representation exists.  Finally, we would prove that this more general
entropy was given by \eqref{s4}, extended to a sum over the full
spectrum of eigenvalues $\lambda$.  We would also need to analyze
the further subtleties that arise
when $\Delta$ has zero-modes which are not also
zero-modes of $R$.  For a fuller discussion of these points, we refer
the reader to \cite{SorkinS}.


We have arrived at a formulation that, for a gaussian field,
expresses the entropy directly in terms of the pairwise correlation
functions of the theory.  As a rudimentary check of our
framework,
one
can examine the ``0+1 dimensional'' case of a harmonic oscillator at
finite temperature.  In Appendix B, we do so and confirm that the
expected result is obtained.

As we have already mentioned, the more generally defined entropy of a
region \eqref{s4} can in certain cases be interpreted as an entanglement
entropy.  In the next two sections we consider two such examples in flat
2-dimensional spacetime.  However, for a massless field, there exists no
consistent vacuum in two-dimensional Minkowski space, $\Minkowski^2$.
For this reason, we will carry out the calculation of
Section 3
in a larger causal
diamond that serves as
infrared cutoff.  We then need to choose a vacuum
for this larger diamond.
Our choice will be based on a recently proposed distinguished ground
state for a free scalar field theory in a globally hyperbolic region or
spacetime.  This ``ground state'' or ``vacuum'' is called the ``SJ''
(Sorkin-Johnston)
vacuum,
and its definition is reviewed in Appendix~A.

For consistency in speaking of entanglement entropy, it is important that
the global entropy vanish.  When the Wightman function is the SJ one
$W_{SJ}$, the entropy does in fact vanish, for the eigenvalues in
\eqref{s4} are by construction either $\lambda=1$ or $\lambda=0$.  Each
term in the sum \eqref{s4} is therefore zero.  This outcome was to be
expected, since the SJ vacuum is a pure state.

\section{Entanglement Entropy I: Small Diamond in Big Diamond}
\begin{figure}[t]
\begin{center}
\includegraphics[width=0.6\textwidth]{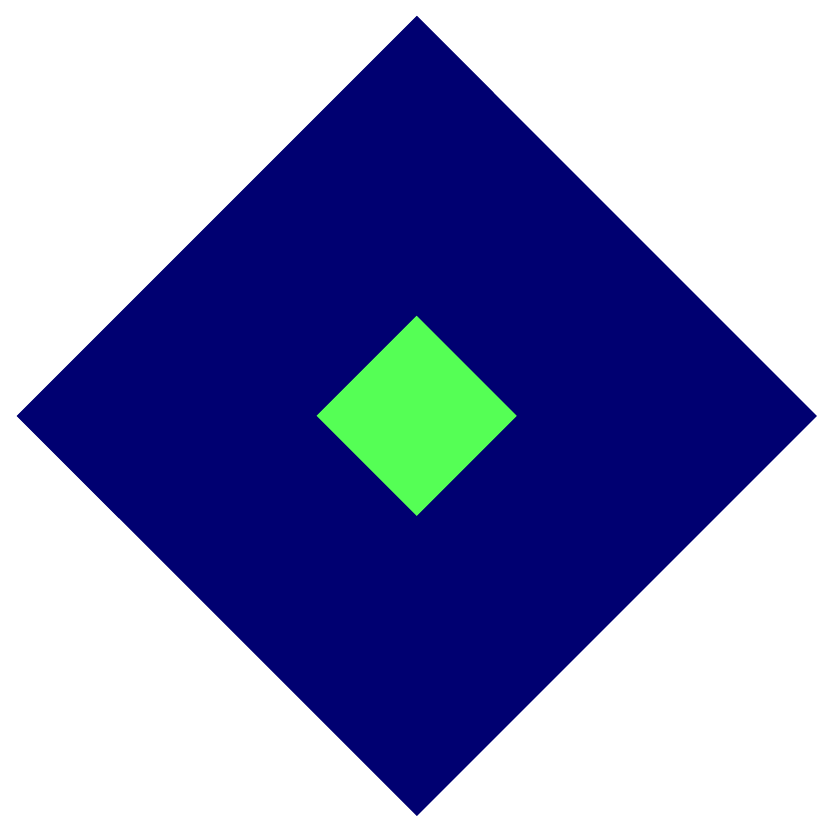}
\caption{A diamond region within a diamond spacetime.}
\label{2d}
\end{center}
\end{figure}

%
We apply the formalism described in the previous section to compute the
entanglement entropy of a causal diamond
embedded in a larger 1+1 dimensional
causal diamond spacetime,
as shown in Figure~\ref{2d}.
A causal diamond
(also called order-interval or Alexandrov neighborhood)
is the intersection of the
future
of a point $p$ with the
past of a point $q \succ p$.
As is evident in the figure,
each diamond is the domain of dependence of the 1d interval that is its
``waist'' or ``diameter''.
Thus our result for the entropy of the smaller diamond should be
compared with the CFT-results for the entanglement entropy between a
shorter line-segment and a longer one containing it.

Usually periodic boundary conditions are imposed in the CFT calculations,
and with this choice,
the entanglement entropy for a massless scalar field
has been found to take the asymptotic form for $a\to0$
\cite{Cardy1, Cardy2} (see also \cite{Ryu}),
\be
    S \sim \frac{1}{3} \ln[\frac{\tilde{L}}{\pi a}\sin(\frac{\pi\tilde{\ell}}{\tilde{L}})] + c \ ,
\label{cs}
\ee
where $a$ is a UV cutoff,
$\tilde{\ell}$ is the length of the shorter interval,
$\tilde{L}$ is the length of the longer interval,
and $c$ is a non-universal constant.\footnote
{The entropy \eqref{cs} is supposed to be defined within an overall
vacuum state, which however doesn't quite exist because a massless
scalar field
on a circle (spacetime cylinder) has a zero-mode, which acts
like a free particle and as such possesses no normalizeable ground
state.  One can take a limit in which its energy goes to zero, however,
and in this limit its contribution to the entanglement entropy seems to
diverge logarithmically.  The entanglement entropy would then
be infinite, even with a UV cutoff.  Presumably the CFT formulas
have in mind regulating the zero-mode, by a small mass or otherwise, and
holding the regulator fixed while sending the UV cutoff $a$ to zero.}
In the limit that the smaller interval is much shorter than the larger
one ($\frac{\tilde{\ell}}{\tilde{L}}\rightarrow 0$),
the entropy reduces to
\be
   S \sim \frac{1}{3} \ln \left[\frac{\tilde{\ell}}{a} \right] + c \ .
\label{csl}
\ee
In this limit, $S$ depends only on the length of the smaller interval
and the UV cutoff of the theory.
For the massive theory, one would expect $1/m$ to play the role of IR
scale, in which case the entropy would take the form
(cf. \cite{Cardy1,Cardy2}),
\be
   S \sim -\frac{1}{3} \ln[ma]
\label{csm}
\ee

We now present a numerical calculation for the massless
scalar field
in
the continuum.
In setting it up
we will borrow freely from reference \cite{Yas}, starting with
the forms of $i\Delta$ and $W$ for our model.
In Minkowski lightcone coordinates
$u=\frac{t+x}{\sqrt{2}}$ and
$v=\frac{t-x}{\sqrt{2}}$,
the Pauli-Jordan function is given by
\be
   \Delta(u,v;u',v') = \frac{-1}{2} [ \theta (u-u')+\theta (v-v')-1] \ .
 \label{pjd}
\ee
For $W$, we will use the asymptotic form of $W_{SJ}$ for a large causal
diamond when the spacetime points of interest lie far from the corners of
the diamond.  In \cite{Yas} this was found to be
\be
  {W}_{\mathrm{centre}}(u,v;u',v')
   = -\frac{1}{4\pi} \Log|\Delta u \Delta v|
     - \frac{i}{4}\text{sgn}(\Delta u+\Delta v)\theta(\Delta u\Delta v)
     -\frac1{2\pi}\Log\frac{\pi}{4{L}}
     +\epsilon_{\mathrm{centre}}
     + \mathcal O\left(\frac{\delta}{{L}}\right),
\label{eq:SJtpcentrecomplete2}
\ee
where $\epsilon_{\mathrm{centre}}\approx -0.063$ and $\delta$
collectively denotes the coordinate differences $u-u',v-v',u-v',v-u'$.
Here, $L=\tilde{L}/\sqrt{8}$ is the ``half side length'' of the larger
diamond.
(It will be convenient to work with ${L}$ and its analog ${\ell}$ for the
smaller diamond, rather than with the diameters $\tilde{\ell}$ and $\tilde{L}$
($\tilde{\ell}=2\sqrt{2}{\ell}$ and $\tilde{L}=2\sqrt{2}{L}$).
We will also set ${\ell}\equiv1$ and choose then ${L}=100$  so that the
$\ell/L=0.01$ and we are in a regime where \eqref{csl} applies.)

Within the smaller diamond the
 $\mathcal O\left(\frac{\delta}{{L}}\right)$ correction in
\eqref{eq:SJtpcentrecomplete2} will be negligible, and we can write the
remainder more simply as
\be
  W(u,v; u',v') =
  \lim_{ \epsilon \rightarrow 0^{+}}
   (-\frac{1}{4\pi } \Log\left[-\mu^2(\Delta u-i\epsilon)(\Delta v -i\epsilon)\right])
  \label{Wsimp}
\ee
where
$\mu=(\pi/4L)e^{-2\pi\epsilon_{\mathrm{centre}}}$
is the IR scale of the large diamond.  As long as the small diamond is
much smaller than the large one, this approximation should be adequate.
In our calculation we will use \eqref{Wsimp} for $W$,
with $\mu$ taken specifically to be $\mu=0.0116681$.

We should note here that although the construction of $W_{SJ}$ for a
causal diamond is completely well-defined, it has no finite limit as the
large diamond goes to infinity.  Indeed a self-consistent Minkowski
vacuum state $|0_M\rangle$ does not exist.  If we try to define a vacuum
in the usual way as the state annihilated by the operator coefficients
of the positive frequency modes in the expansion of the field operator
$\hat{\phi}(t,x)$, then we encounter an infrared divergence.  We can
remove the divergence by introducing a long wavelength cutoff into the
integral for the Wightman function $W$, but the result is unphysical because
it fails to be positive semidefinite as a quadratic form.
Nevertheless, the resulting expression matches the general form
\eqref{Wsimp} that we obtained as a local approximation to the SJ vacuum
of the large diamond.
In this sense, we can think of \eqref{Wsimp} as an approximate Minkowski
vacuum which is valid for separations $\Delta t$ and $\Delta{x}$ that
are small compared to the IR scale $\mu$.

Returning to our calculation,
we want to solve
\be
    W v=i\lambda\Delta v    \label{wds}
\ee
subject to
\be
    \Delta v\neq 0 \ .  \label{nk}
\ee
To that end we will represent $W$ and $\Delta$ as matrices, using the basis
that diagonalizes $i\Delta$, and which consists of two
families of eigenfunctions:\footnote%
{Thanks to \eqref{nk} we need only consider functions orthogonal to the
  kernel of $\Delta$, all of which consist of solutions to the wave
  equation.  If one wanted to expand arbitrary $L^2$ functions, one
  would need to supplement the solutions,
  \eqref{eq:SJfunctions3} and
  \eqref{eq:SJfunctions4},
 with a basis for $\ker\Delta$.}
\begin{align}
  f_k(u,v) &:= e^{-iku} - e^{-i k v}, & &\textrm{with } k = \frac{n \pi}{{\ell}}, \; n =\pm 1,\pm 2, \ldots\label{eq:SJfunctions3}\\
  g_k(u,v) &:= e^{-iku} + e^{-i k v} - 2 \cos(k {\ell}), & &\textrm{with } k\in\mathcal{K}, \label{eq:SJfunctions4}\end{align}
where
$\mathcal{K}=\left\{k\in\mathbb{R}\,|\,\tan(k{\ell})=2k{\ell} \ \textrm{and} \ k\neq0\right\}$.
The eigenvalues are ${\lambda}_k={\ell}/k$.
The $L^2$-norms are $||{f}_{\small{k}}||^{2}=8{{\ell}}^{2}$
and
$||{g}_{\small{k}}||^{2}=8{{\ell}}^{2}-16{{\ell}}^{2}{\cos}^{2}(k{\ell})$.

Before actually emabarking on the numerics, however, we need to decide
on a cutoff.  As we have been emphasizing, it will necessarily have a
spacetime character as opposed to the purely spatial one seen, for
example, in a lattice of carbon atoms.  A discrete theory provides its
own cutoff, but here in the continuum a naive lattice cutoff would be
inconvenient and possibly inappropriate.  Instead we simply truncate the
matrices representing $W$ and $\Delta$ by retaining only a finite number
of eigenfunctions $f_k$ and $g_k$ up to a maximum value $k_{max}$ of
$k$.  Finally, in comparing our results with \eqref{csl}, we need to
translate our cutoff into a purely spatial one $a$.  It is not certain that
such a correspondence is always possible, but in this case we are
expanding solutions of the wave equation, which in turn are in
one-to-one correspondence with initial data specified on the spatial
diameter of the causal diamond.  With the modes we have retained, we can
expand initial data of wavelengths longer than
$\lambda_{min}\sim 1/k_{max}$ (or $2\sqrt{2}\pi/k_{max}$ if one were
trying to be more precise).
It is therefore natural to equate $a$ to $1/k_{max}$, and this is what we do in
the comparisons below.

\begin{figure}[t]
\begin{center}
\includegraphics[width=0.8\textwidth]{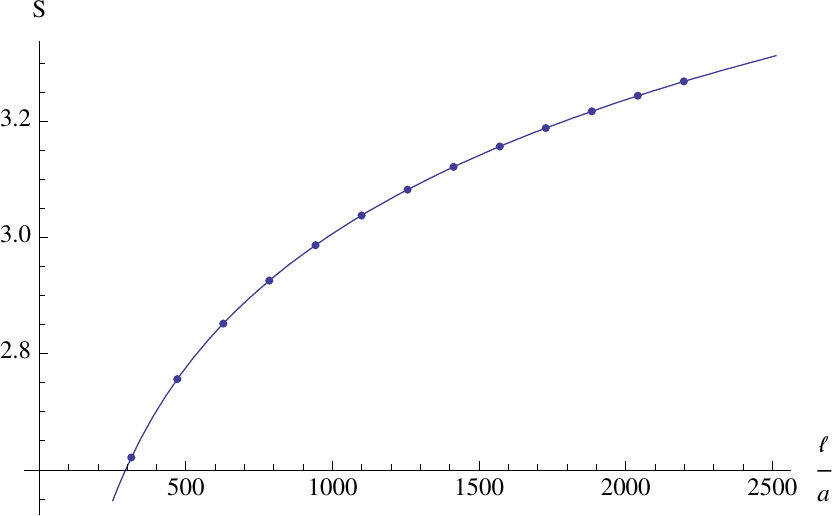}
\caption{The entanglement entropy $S$ versus ${\ell}/{a}$.
    Data points represent calculated values of \eqref{s4}. }
\label{sfg}
\end{center}
\end{figure}

In our basis, the integral-kernel $i\Delta$ is diagonal, so its
representation is trivial, but for $W$, we must compute
$\langle f_k|W|f_{k'}\rangle$ and $\langle g_k|W|g_{k'}\rangle$,
which we did numerically.\footnote%
{~We performed the calculations in this section and the next using Mathematica 9.0.}
The terms $\langle f_k|W|g_{k'}\rangle$ vanish, making $W$ block
diagonal in this basis,
so we can treat each block separately in solving \eqref{wds}.
Summing over the resulting eigenvalues $\lambda$, we obtain the entropy
associated to each block.
Each block contributed to the entropy roughly equally, with the $g$
block making a slightly greater contribution.
Adding the two contributions, we obtain the total entropy.
In the calculations reported here, all of the eigenvalues
obtained from \eqref{wds}-\eqref{nk} were order-one
numbers of absolute value below $3$, with all but a handful of the
eigenvalue-pairs being very close to the values one and zero.
(As required for consistency we did not encounter any functions in the
kernel of $i\Delta$.)
The resulting entropies are plotted in Figure \ref{sfg},
as a function of  $\frac{{\ell}}{a}$.
As seen in the plot, the obtained values of $S$ are fit almost perfectly
by the curve
\be
      S = b \ln\left[ \frac{{\ell}}{a}\right] + c
\ee
with
$b= 0.33277 $ and 
$c=0.70782$.
Thus, the entropies obtained from our ``spacetime formulation''
closely match the asymptotic form \eqref{csl}.
%

\section{Entanglement Entropy II: Diamond in Halfspace}
%
In the previous section, we calculated the entropy of a causal diamond
in $\Minkowski^2$, or more accurately a diamond embedded in the center
of a much larger diamond.
In this section,
we do the analogous calculation for a causal diamond
embedded in
(and touching the boundary of)
a spacetime equal to
the right half of $\Minkowski^2$.
In terms of subregions of a spacelike hypersurface
(or rather hyposurface)
we are here computing
entanglement entropy for
a 1d interval at one end of a semi-infinite line,
while
in the previous section our interval was centered within a much larger
but still finite interval.

Compared to full $\Minkowski^2$, the half-space has the advantage that
it admits a true minimum energy state or vacuum, relieving us of the
need for an infrared cutoff (other than the cutoff always imposed by
the electronic computer.)
On the other hand, the presence of a boundary requires that we choose a
boundary condition.  For the free massless scalar field which we
consider, we will require the field to vanish at the boundary (``Dirichlet
condition''), and our vacuum will be the ground state with respect to this
condition.  Of course the calculation itself cares nothing
about the boundary, except indirectly insofar as it influences the Wightman
function.

\begin{figure}[h!]
\begin{center}
\includegraphics[width=0.9\textwidth]{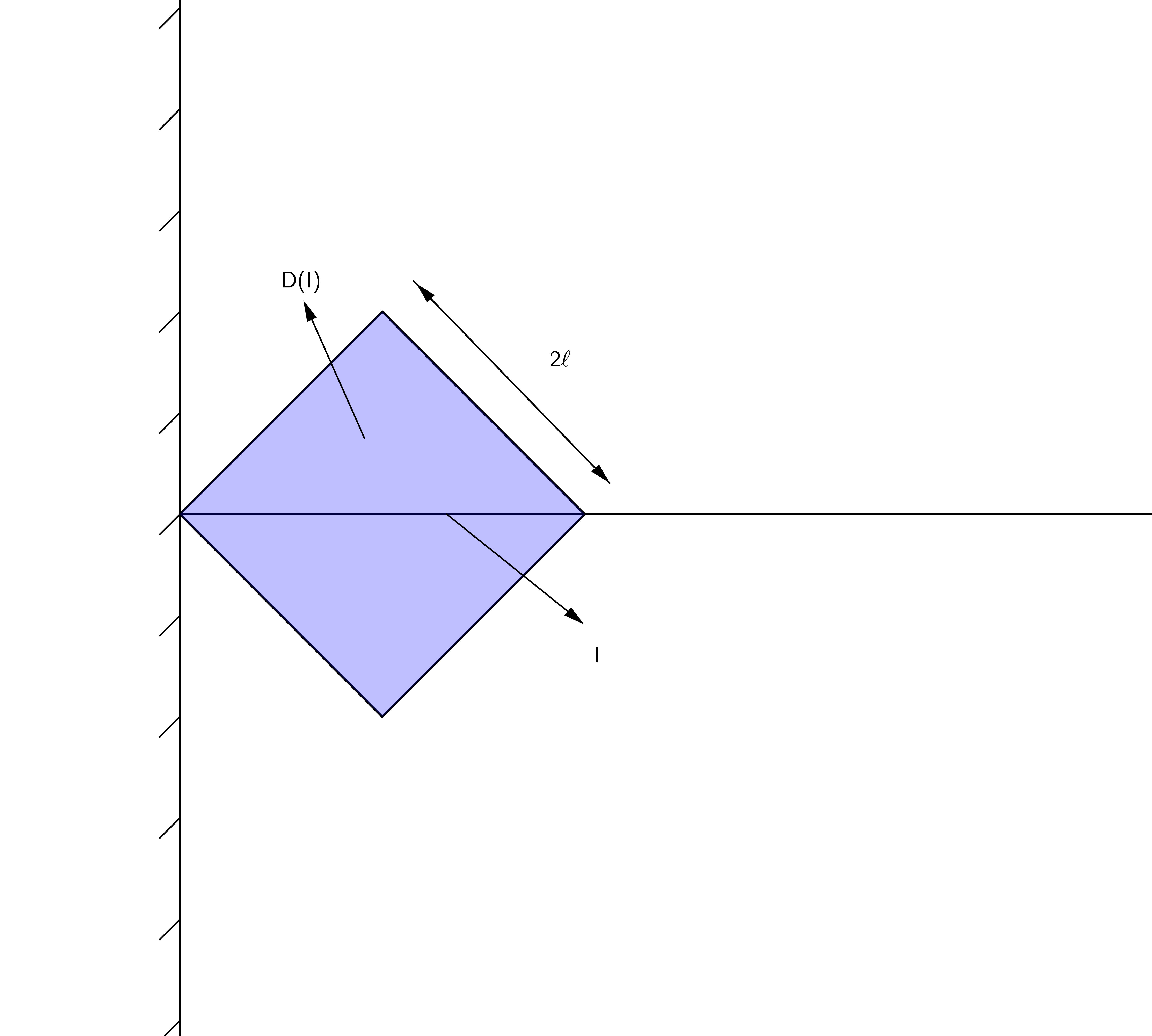}
\caption{Spacetime diagram of the region of interest in this section.}
\label{semiline}
\end{center}
\end{figure}

Our spacetime will
comprise the subset of $\Minkowski^2$ defined by
$t\in(-\infty,+\infty)$
and $x\in[0,+\infty)$.
The diamond whose entropy we seek will be the one shown in Figure~\ref{semiline},
its diameter being the interval
$I=\{(t,x)|~t=0, \   x\in[0,2\sqrt{2}\ell]\}$.
As before, we will compute the entropy
of a free massless scalar field $\phi(x)$.
Under the boundary condition, $\phi(x=0,t)=0$, the solutions to the wave
equation are
\be
  \mathcal{U}_k(x,t)=\sqrt{\frac{2}{|k|}}e^{-i k t}\sin(k x),
\ee
resulting in the following two point function for the vacuum state
\be
  W(x,t;x',t')=\frac{1}{2 \pi}\int_{-\infty}^{+\infty}\frac{dk}{|k|} e^{-i |k|(t-t')}\sin(k x) \sin(kx').
\ee
One can explicitly check that $W(X,X')-W(X',X)=i\Delta(X,X')$
for $X,X'\in D(I)$, with $\Delta$ given by \eqref{pjd}.

\begin{figure}[t]
\begin{center}
\includegraphics[width=\textwidth]{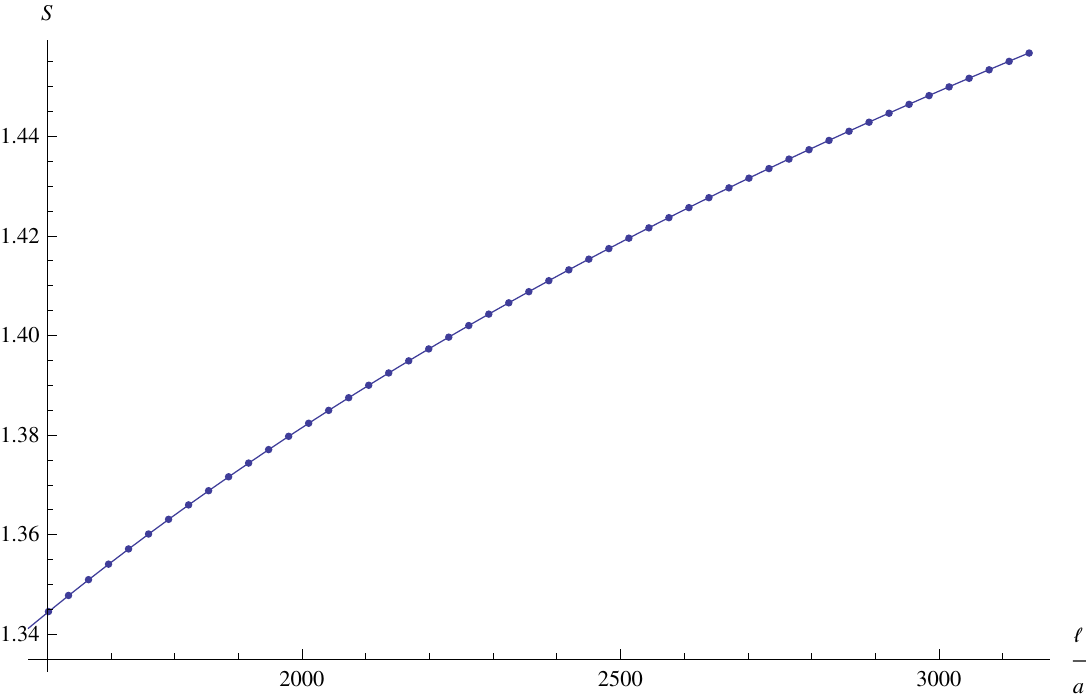}
\caption{The entanglement entropy $S$ versus the UV cutoff $a$ for the case of the halfspace.}
\label{Sfgsemi}
\end{center}
\end{figure}

As before, we need to solve the eigenvalue problem,
\be
  Wv=i \lambda  \Delta v \ ,
\ee
where the integration is over the
shaded region in
Figure~\ref{semiline}.
(This would be the full domain of dependence of $I$, were we in all of
$\Minkowski^2$, but given the boundary it is only a subset thereof.
The resulting loss of information for numerical purposes is compensated
by the convenience of working within a rectangular shaped region.)
Since $i\Delta$ is exactly the same operator as before, we will use the
same basis-functions as in the previous section.
(Notice however that the centre of the causal diamond is no longer the
centre of the coordinate system.)
The
representation of $\Delta$ is trivial in this basis, while
representing $W$
requires us to compute
$\langle f_k|W|f_{k'}\rangle$,
$\langle f_k|W|g_{k'}\rangle$,
and $\langle g_k|W|g_{k'}\rangle$.
Calculating these inner products and introducing the same UV cut-off,
$a$, we are able to evaluate the entanglement entropy.  The result, as
shown in Figure \ref{Sfgsemi},
is fit almost perfectly by
\be
  S = \frac{1}{6}\ln\left[\frac{\ell}{a}\right]+c \ ,
\ee
with $c=0.11465$.
This again agrees with the CFT asymptotic form.\footnote%
{The coefficient is $1/6$ rather than $1/3$ because the entanglement
 concerns only one of the two boundaries of the smaller interval.}

\pagebreak

\section{Acknowledgements}
We would like to thank
Niayesh Afshordi,
Siavash Aslanbeigi,
Achim Kempf,
and
Rob Myers
for discussions.
This research was supported in part by Perimeter Institute for
Theoretical Physics.  Research at Perimeter Institute is supported by
the Government of Canada through Industry Canada and by the Province of
Ontario through the Ministry of Research and Innovation.
This research was supported in part by NSERC through grant
RGPIN-418709-2012.

\pagebreak

\appendix
\section{The S-J Ground State}
We review the SJ proposal for the ground state of a free scalar field
in a $d$-dimensional globally hyperbolic
spacetime
$\mathcal M$ with metric $g_{\mu\nu}\,$.
\cite{Sorkin:2011pn, Johnston:2009fr}.
The starting point is the Pauli-Jordan function $\Delta  (X,X')$, where
$X, X'$ denote spacetime points. It is defined by
\be
\Delta(X,X'):=G_R(X,X')-G_R(X',X)
\label{eq:pjdef}
\ee
where $G_R$ is the retarded Green function that satisfies $G_R(X,X')=0$
unless $X'\prec X$, meaning that $X'$ is to the causal past of $X$.  The
integral-kernel $\Delta$ is real and antisymmetric.  It is related to
the commutator by:
\be
  i \Delta  (X,X') = [ \hat\phi(X), \hat\phi(X')  ]   \ .
  \label{eq:paulijordancommutator}
\ee

The Wightman function, or two-point function of a state $|0\rangle$, is
\be
   W_0(X, X') = \langle 0| \hat\phi(X)\hat\phi(X')|0\rangle \,.
\ee
The SJ vacuum is defined through its Wightman function by
the three conditions \cite{Yas}:
\begin{enumerate}
\item \emph{commutator: }  $i\Delta(X,X') = W(X,X')-W^*(X,X')$
\item \emph{positivity: } $\int_{\cal{M}} dV \int_{\cal{M}} dV' f^*(X) W(X, X') f(X') \ge 0$
\item \emph{orthogonal supports: }
  $ \int_{\cal{M}} dV'  \; W(X, X') \, W(X',X'')^*  = 0 $,
\end{enumerate}
where $\int dV = \int d^d X \sqrt{-g(X)}$.

These conditions have the meaning that $W$ is the positive part of $i\Delta$,
thought of as an operator on the Hilbert space of square integrable
functions ${L}^{2}( \mathcal M,dV)$ \cite{Sorkin:2011pn}.
This allows us to describe a direct construction of $W$
from the Pauli-Jordan function \cite{Johnston:2009fr, Johnston:2010su, Siavash}.
The distribution $i\Delta(X,X')$ defines the kernel of a (Hermitian) integral operator
(which we may call the \emph{Pauli-Jordan operator} $i\Delta$)
on ${L}^{2}( \mathcal M,dV)$.  The inner product on this space is
\be
\langle f, g \rangle:=\int_{\mathcal M} dV f(X)^*g(X),
\ee
where $dV=\sqrt{-g(X)}d^d X$ is the invariant volume-element
on $\mathcal M$.
The action of $i\Delta:f\mapsto i\Delta f $ is then given by
\be
(i\Delta f)(X) := \int_{\mathcal M }^{ } dV' i\Delta(X,X') f(X').
\label{eq:defpjfunctional}
\ee
When this operator is self-adjoint, it admits a unique spectral
decomposition.
Whether or not this (or some suitable generalization of it) is the case
depends on the functional form of the kernel $i\Delta(X,X')$ and on the
geometry of $\mathcal M$.
For the massless scalar field on a bounded region of
Minkowski space, such as the finite causal diamond considered in this paper, $i\Delta$ is indeed a self-adjoint operator, since the
kernel $i\Delta(X,X')$ is Hermitian and bounded. In the following we assume that
we can expand
$i\Delta$
in terms of its eigenfunctions.

Noting that the kernel $i\Delta$ is skew-symmetric, we find that the
eigenfunctions in the image of $i\Delta$ come in complex conjugate pairs
$T^+_q $ and $T^-_q $ with real eigenvalues $\pm \lambda_q $:
\be
(i\Delta T^\pm_q )(X)=\pm\lambda_q  T^\pm_q (X),
\ee
where $T^-_q (X)=[T^+_q (X)]^*$ and $\lambda_q >0$.
Now by the definition of $i\Delta$, these eigenfunctions must be
solutions to the homogeneous Klein-Gordon equation.
If they are $L^2$-normalised
so that $||T^+_q ||^2:=\langle T^+_q ,T^+_q \rangle=1$,\footnote%
{~When the spacetime region has infinite volume, this must be replaced by a
  delta-function normalisation $\langle T^+_q ,T^+_{q'}\rangle=\delta(q -q ')$.}
then the spectral decomposition of
the Pauli-Jordan
operator
implies that
its
 kernel can be written as
\be
  i \Delta  (X,X' ) =
  \sum_{   q }^{ }\lambda_{q } T^+_{q }(X)T^+_{q }(X')^*-\sum_q \lambda_{q } T^-_q (X)T^-_q (X')^\star \ .
  \label{eq:SJpaulijordan}
\ee
We
construct the SJ two-point function ${W}_{SJ}(X,X')$ by
restricting~\eqref{eq:SJpaulijordan} to its positive part:
\be
   {W}_{SJ}(X,X') :=
   \sum_{ q }^{ } \lambda _{q }T^{+}_{q}(X) T^+_{q }(X')^\star =
   \sum_{ q }^{ } \mathcal T_{q }(X) \mathcal T_{q }(X')^\star
  \label{eq:SJwightman}
\ee
where ${\mathcal{T}}_q (X):= T^+_q(X)\sqrt{\lambda_q}$.

\section{A Simple Illustration: Thermal Entropy of a Harmonic Oscillator}

Let us apply the new entropy formula to the harmonic oscillator in
one dimension.
First we compute the entropy using the standard thermodynamic relation
\be
  S=\frac{\partial}{\partial T}( T\ln Z)
\label{hos}
\ee
For the harmonic oscillator, with energies $E_n=(n+\frac{1}{2})\omega$,
the partition function is
\be
    Z=\frac{e^{-\frac{\beta\omega}{2}}}{1-e^{-\beta\omega}}
   \label{hoz}
\ee
where as usual $\beta=1/k_{\small{B}} T$ and we set $k_{\small{B}}\equiv 1$.
From \eqref{hoz} and \eqref{hos}  we obtain the entropy as
\be
  S = -\ln[1-e^{-\beta\omega}] + \frac {e^{-\frac{\beta\omega}{2}}} {1-e^{-\beta\omega}} \ .
     \label{ssho}
\ee

Now we turn to computing the entropy using the new formula.
With the field-operator $\hat\phi(t)$ identifed as the oscillator's
position-operator $\hat{q}(t)$ in the Heisenberg picture
(and with the oscillator's mass set to unity or absorbed into $q$),
we have for the commutator
\be
   i\Delta(t,t') = \frac{1}{2\omega}\left(e^{-i\omega (t-t')}-e^{i\omega (t-t')}\right),
   \label{B4}
\ee
and for the thermal Wightman function,
\be
   W(t,t') = \frac{1}{2\omega}\left(\frac{e^{-i\omega(t-t')}+e^{i\omega(t-t')}}{e^{\beta \omega}-1}+e^{-i\omega(t-t')}\right).
\ee
In order to find the entropy, we need to solve the eigenvalue
equation \eqref{gee}, which in the present context says
\be \label{1}
    \int_L W(t,t')f(t')dt' = \lambda\int_L i\Delta(t,t')f(t')dt',
\ee
the integration being over the interval $L$.

Since $f$ is required to belong to the image of $\Delta$, it must be
a linear combination of $e^{\pm i \omega t}$, as is evident
from \eqref{B4}.  Writing then
\be
A_{\pm} \equiv \int_L e^{\pm i \omega t}f(t) dt,
\ee
we learn from (\ref{1}) that
\be
   e^{-i\omega t}\left(\frac{e^{\beta\omega}}{e^{\beta\omega}-1} - \lambda\right) A_+
   = -e^{i\omega t}\left(\lambda+\frac{1}{e^{\beta\omega}-1}\right) A_- \ .
   \label{B8}
\ee
As this must hold for any value of $t\in L$, the coefficients
of $e^{i\omega t}$ and $e^{-i\omega t}$ must be zero.
Hence, we obtain two eigenvalues, each with multiplicity one,
\begin{eqnarray}
\lambda&=&\frac{e^{\beta\omega}}{e^{\beta\omega}-1} \quad  \mbox{from} \ A_-=0,~ A_+\neq 0 ~~~\mbox{and}\\
\lambda&=&-\frac{1}{e^{\beta\omega}-1}  \quad  \mbox{from} \  ~ A_+=0,~ A_-\neq 0\ .
\end{eqnarray}
%
Substituting these two eigenvalues into
\be
   S = \sum \lambda \, \ln|\lambda|
\ee
we obtain \eqref{ssho},
the desired result.


\section{Symmetry of Entanglement from the Basic Formula \eqref{s4} }
\label{Appendix C}
%
Given a ``bipartite'' quantum system in an overall pure state, one knows
that the entropies of the separate subsystems are necessarily equal.
Here,
after recalling the proof of this fact from the
existence of Schmidt decompositions, we show how the same equality follows
directly from our basic formula \eqref{s4}.

Let some Cauchy surface be divided into a subregion $A$ and the
complementary subregion $B$.  Let
\be
  \rho_A = \Tr_B \rho
\ee
be the reduced density matrix for region $A$, and let
\be
        S_A = - \Tr\rho_A\ln\rho_A
\ee
be the entropy of this reduced density matrix.
That one characterizes $S_A$ as simply ``the entanglement entropy'',
when the overall state of the field is pure, owes its consistency to the
fact that it doesn't matter which subregion one looks at: $S_A=S_B$.

Of course this equality is formal, since it relates two infinite
quantities.  However, in a finite-dimensional Hilbert space, it holds
rigorously thanks to the  Schmidt decomposition theorem:
For any vector $\psi_{AB}$ in a tensor product Hilbert space, there
exist orthonormal sets $\{\psi_A\}$ and $\{\psi_B\}$ such that
\be
  \psi_{AB}=\sum_n\lambda_n^{1/2} \psi_A^{(n)} \psi_B^{(n)}
\label{sdt}
\ee
where $\lambda_n^{1/2}>0$ and $\sum_n\lambda_n=1$.   (This holds even if
the Hilbert spaces $\mathcal{H}_A$ and $\mathcal{H}_B$ have different
dimensions, in which case the index $n$ in \eqref{sdt} cannot exceed the
smaller of the two dimensions.)  Now
\be
  \rho_A = \Tr_B \psi^{\phantom{\dagger}}_{AB} \psi^{\dagger}_{AB}
         = \sum_n\lambda_n \psi_A^{(n)} \psi^{(n)\dagger}_A
\ee
and
\be
  \rho_B = \Tr_A \psi_{AB} \psi^{\dagger}_{AB}
         = \sum_n\lambda_n \psi_B^{(n)} \psi^{ (n)\dagger}_B
\ee
Therefore,
since $\rho_A$ and $\rho_B$ share the same nonzero
eigenvalues $\lambda_n$,
it follows that $S_A=S_B$.

\begin{figure}[t]
  \begin{center}
  \includegraphics[width=.6\textwidth]{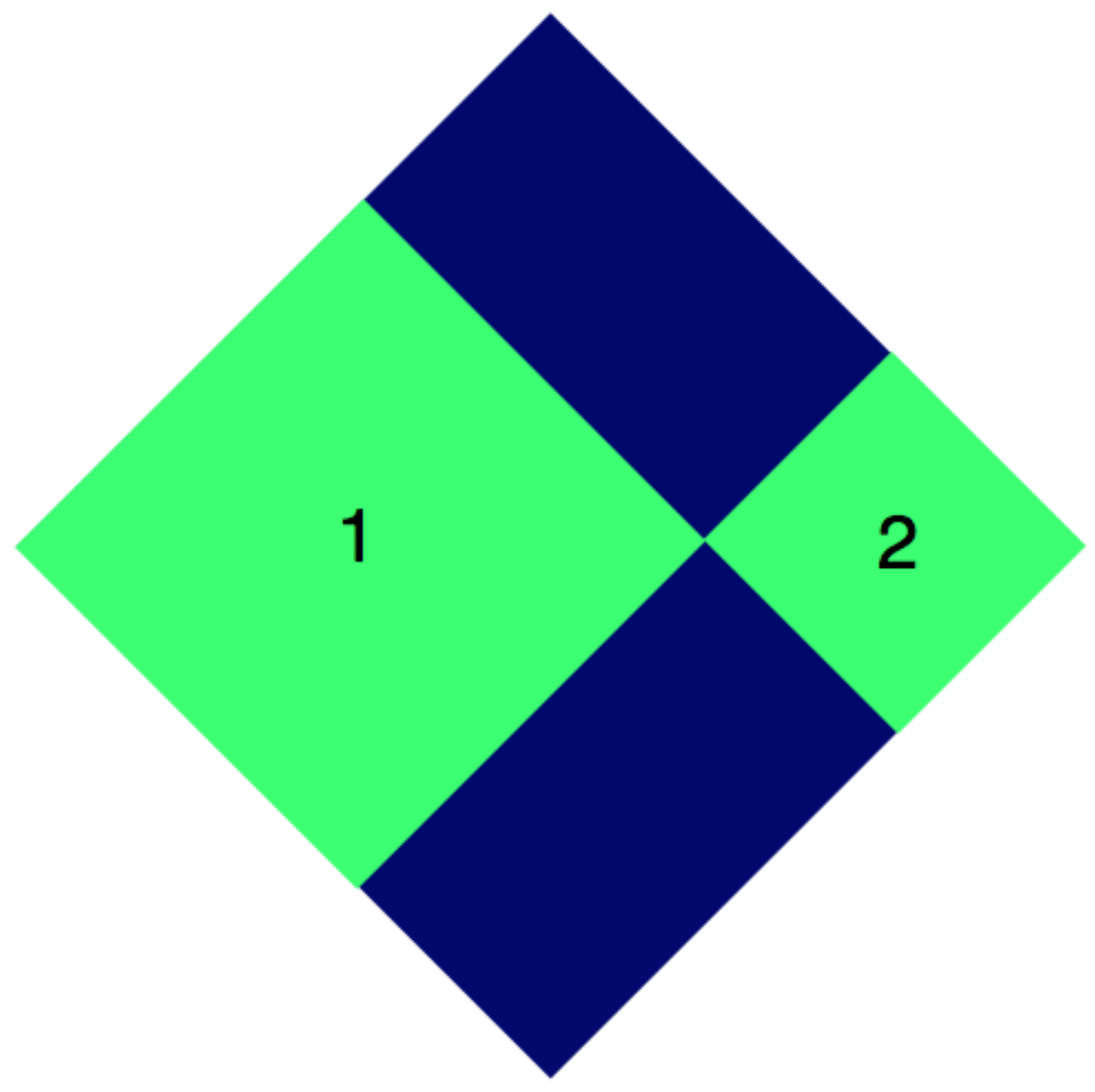}
  \caption{Division of a spacetime by two globally hyperbolic regions, $1$ and $2$.}
  \label{ab}
  \end{center}
\end{figure}

We now  prove this basic property of entanglement, using
the new formulation.
Let us divide the spacetime
as shown in Figure \ref{ab},
where
$1$ and $2$ are
causally disjoint globally hyperbolic regions
whose union contains the whole spacetime
in its ``domain of dependence''
(the union contains a Cauchy surface for the full spacetime ).
Restricting, now,
the two-point functions,
$W$ and $\Delta$,
to the union of the two regions,
we can
write them in block-matrix form as follows,
where the zeroes in $\Delta$ express the vanishing of the commutator at
spatial separations.
\[\Delta= \left( \begin{array}{cc}
  \Delta_{11} & 0 \\
  0 & \Delta_{22} \\\end{array} \right)\]
%
%
\[W= \left( \begin{array}{cc}
  W_{11} & W_{12} \\
  W_{21} & W_{22} \\
\end{array} \right)\]

For simplicity,
let us now pretend that $\Delta$ is invertible, and define
$M\equiv\Delta^{-1} W$.
Because the overall state is pure by assumption,
we know that $M$ must have eigenvalues
$0$ and $1$ and no others.
(Otherwise the entropy would not vanish.)
As an operator equation this says $M^2=M$,
which in turn yields when written out
fully the two equations,
\be
  (\Delta_{11}^{-1} W_{11})^2-\Delta_{11}^{-1} W_{11}=\Delta_{11}^{-1} W_{12}\Delta_{22}^{-1} W_{21}
\ee
\be
  (\Delta_{22}^{-1} W_{22})^2-\Delta_{22}^{-1} W_{22}=\Delta_{22}^{-1} W_{21}\Delta_{11}^{-1} W_{12}
\ee
This pair of equations has the form,
$Q_{1}^2-Q_{1}=RS$ and
$Q_{2}^2-Q_{2}=SR$,
where
$Q_1=\Delta_{11}^{-1} W_{11}$ and
$Q_2=\Delta_{22}^{-1} W_{22}$.
Using now the general fact\footnote%
{The nonzero spectrum of a finite-dimensional matrix $M$
 can be deduced directly from the traces of its powers, $\Tr(M^n)$.
 (More precisely, one can deduce the multiset of its eigenvalues.)  But
 cyclicity of the trace implies that for all $n$,
 $\Tr[(RS)^n]=\Tr[(SR)^n]$.  Hence the matrix products $M=RS$ and $M=SR$
 share the same nonzero spectrum.  Notice that in our situation, the
 matrices $R$ and $S$ are not necessarily square because regions 1 and 2
 are not necessarily of the same size.}
that
the
nonzero
spectrum of the product of two matrices
is independent of the order in which the product is taken
(this includes the multiplicity of the eigenvalues),
we can conclude that
$RS=Q_{1}^2-Q_{1}$
and $SR=Q_{2}^2-Q_{2}$
share the same nonzero eigenvalues:
\be
  \lambda_1^2-\lambda_1 = \lambda_2^2-\lambda_2\indent
  \implies
  \indent(\lambda_1-\lambda_2)(\lambda_1+\lambda_2-1)=0
\ee
For the eigenvalues $\lambda$ that figure in equation \eqref{s4}, this means
\be
   \begin{cases}
    \lambda_1=\lambda_2\\
    \lambda_1^\prime = 1 - \lambda_2 = \lambda_2^\prime
  \end{cases}
\ee
where the last line follows from
the fact that
the eigenvalues of
$\Delta_i^{-1}W_i$ come in pairs,
$\lambda$ and $\lambda'=1-\lambda$.
Therefore
$\Delta_{11}^{-1} W_{11}$ and
$\Delta_{22}^{-1} W_{22}$
have the same
(nonzero)
spectrum,
and $S_1=S_2$.

\newpage
\bibliography{EEntropybib}
\bibliographystyle{jhep}

\end{document}